\documentclass[reprint,fleqn,10pt]{wlscirep}
\usepackage[utf8]{inputenc}
\usepackage[T1]{fontenc}
\usepackage{bm}
\usepackage{multicol}
\usepackage{subcaption}
\usepackage{enumitem}
\usepackage{tcolorbox}
\newcommand{\bb}{{\bm b}}
\newcommand{\bj}{{\bm j}}
\newcommand{\bu}{{\bm u}}
\newcommand{\bx}{{\bm x}}

\newcommand{\bomega}{{\bm \omega}}
\newcommand{\bQ}{{\bm Q}}
\newcommand{\lamb}{{\bm u\times\bm \omega}}
\newcommand{\bnabla}{{\bm \nabla}}
\newcommand{\bxiQdivQ}{{\xi \bm Q(\bm \nabla\cdot \bm Q)}}
\usepackage[switch]{lineno} 
\usepackage{float}
\usepackage{graphicx}
\usepackage{comment}
\usepackage{enumerate}
\usepackage{amsmath,nccmath}
\usepackage{xcolor}
\usepackage{mathrsfs}
\usepackage{dcolumn}

\title{Universal relaxation of turbulent binary fluids}

\author[1]{Nandita Pan}
\author[1,*]{Supratik Banerjee}
\author[1]{Arijit Halder}
\affil[1]{Department of Physics, Indian Institute of Technology, Kanpur, INDIA, 208016}
\affil[*]{e-mail: sbanerjee@iitk.ac.in}

\begin{document}
\twocolumn[
  \begin{@twocolumnfalse}

\begin{abstract}
Upon quenching the forcing, a turbulent system tends to attain the state of stable equilibrium through the process of turbulent relaxation. Such relaxation in binary fluids is of surmount interest for both fundamental science understanding and industrial applications. A systematic investigation of the same has been carried out, for the first time, using direct numerical simulations of Cahn-Hilliard-Navier-Stokes equations. With the help of a thorough scanning, the bulk of each fluid and its interface are found to relax in a different way. However, using the principle of vanishing nonlinear transfer, we propose a convincing, universal pathway of obtaining the turbulent relaxed states for both the bulk and the interface which attain a relaxed state when the turbulent cascades of the inviscid invariants are suppressed. Interestingly, the relaxation of the bulk turns up to be subtly different from the turbulent relaxation of a single hydrodynamic fluid and the interface relaxation is found to follow a Helmholtz-like pressure-balanced condition. 
\end{abstract}

\flushbottom
\maketitle
   \end{@twocolumnfalse}
]
\thispagestyle{empty}

Binary fluids encompass a wide range of two-component systems consisting of  natural oil-water mixtures, cosmetic fluids, active binary suspensions \textit{etc}\cite{Stalidis1990, Erucar2016,tiribocchi2015, Scarbolo2015, Cates2018}. Above a critical temperature ($T_c$), they exist as a homogeneous phase-mixed  emulsion which finds its prolific usage starting from industrial applications (\textit{e.g.,} food, chemical, pharmaceutical products \textit{etc.}) to droplet dynamics in oceanic and atmospheric turbulence \cite{Pascual2021,chan2021,mazzitelli2003turbulent, serizawa1975,Bailey1993, narsimhan2019guidelines,Shenoy2015}. Below $T_c$, binary fluids tend to attain a stable phase separated state (with minimum free energy) through coarsening dynamics or spinodal decomposition  \cite{Bray1995, hohenberg1977, chaikin1995,Koga1993}. Such coarsening can be arrested by the presence of a driven turbulent flow which fragments these separated domains by virtue of its enhanced mixing properties thus leading to a non-equilibrium steady emulsion at large times\cite{berti2005,perlekar2014,perlekar2019, Pan2022,Mukherjee2019}. Upon the quenching of the turbulence drive, the system is expected to relax towards the equilibrium phase-separated state. A thorough understanding of this turbulent relaxation process is therefore pivotal in controlling the state of emulsion in a binary fluid system. 

Albeit largely studied, a pristine characterization of the turbulent relaxed states in neutral fluids and plasmas has been a matter of long-standing debate. While a Beltrami-Taylor type force-free relaxed state has been observed in cosmic plasmas \cite{Chandrasekhar1958, Woltjer1958a,Taylor1974}, a clear signature of Grad-Shafranov type pressure-balanced state has been found in incompressible fluids and plasmas with moderate plasma-$\beta$ \cite{Zhu1995, Sato1996,Kraichnan1988}. 
Although the aligned states are explained using the principle of selective decay, a general theory accounting for the non-vanishing pressure gradient was lacking until recently where a universal mechanism for fluids and plasma relaxation has been proposed based on the suppression of turbulent cascades of inviscid invariants \cite{Banerjee2023}.

Relaxation in binary fluids has been investigated under various situations including the transition of the binary fluid above and below the critical temperature\cite{Koga1993}, the dielectric relaxation of polar binary mixtures\cite{Baba1969}, the long-time glasslike relaxation\cite{Benzi2011}, secondary relaxation in supercooled binary fluids\cite{Harbola2003}, relaxation of optically heated binary colloids etc\cite{Araki2022}. Despite substantial practical importance,  turbulent relaxation in binary fluids has not been explored to date. Interestingly, unlike an ordinary fluid, a binary fluid system is governed by the Cahn-Hilliard-Navier-Stokes (CHNS) equations which collectively describe the evolution of both the bulk of the individual fluids and their mutual interface. When the driving force is switched off, both the bulk and the interfacial regions are expected to go through turbulent relaxation. Our principal objective includes the obtention and complete characterization of such relaxed states. In particular, (i) whether 
such a state will be a force-free or a pressure-balanced state, (ii) whether there exists a universal way to describe the relaxation both in the bulk of each fluid and their interfaces, and finally (iii) whether the bulk of individual fluids relaxes similar to an ordinary hydrodynamic fluid. These questions are fundamental from the perspective of both complex fluid dynamics and industrial applications. 

In this Article, we address these questions in order and show that the turbulent relaxation of a binary fluid is not exactly similar to that of a single hydrodynamic fluid or plasma. Further, the way the bulk relaxes is significantly different from that of the interface. However, the relaxation of both the bulk and the interface can be universally described using the principle of vanishing nonlinear transfer (PVNLT)\cite{Banerjee2023}. Finally, we show that the turbulent relaxation of each component fluid deviates from the relaxation of a single fluid where a pressure balanced relaxation is obtained before the fluid ceases to flow.  
    
\begin{figure*}[ht!]
     \centering
     \begin{subfigure}[t]{\textwidth}
         \centering         \includegraphics[width=0.9\textwidth]{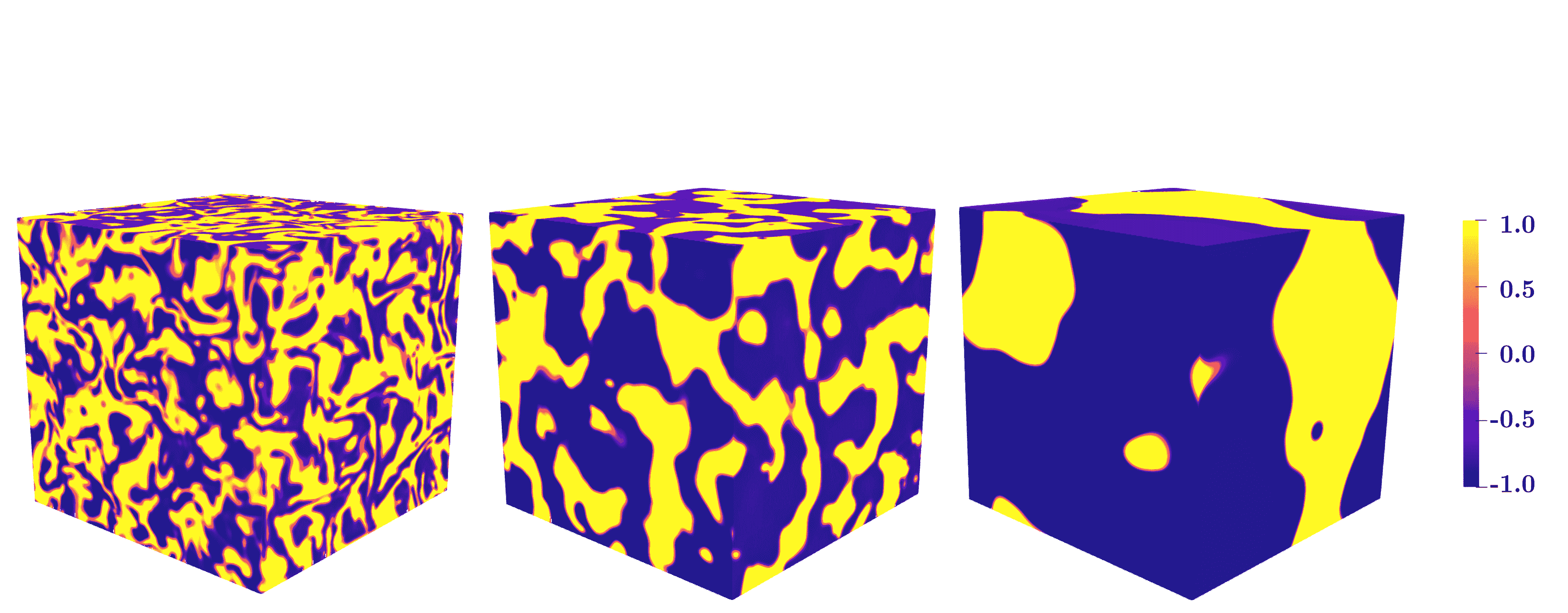}
      \subcaption{}
\label{fig:phi_color}
     \end{subfigure}
     \hfill
     \begin{subfigure}[b]{0.33\textwidth}
         \centering
         \includegraphics[width=\textwidth]{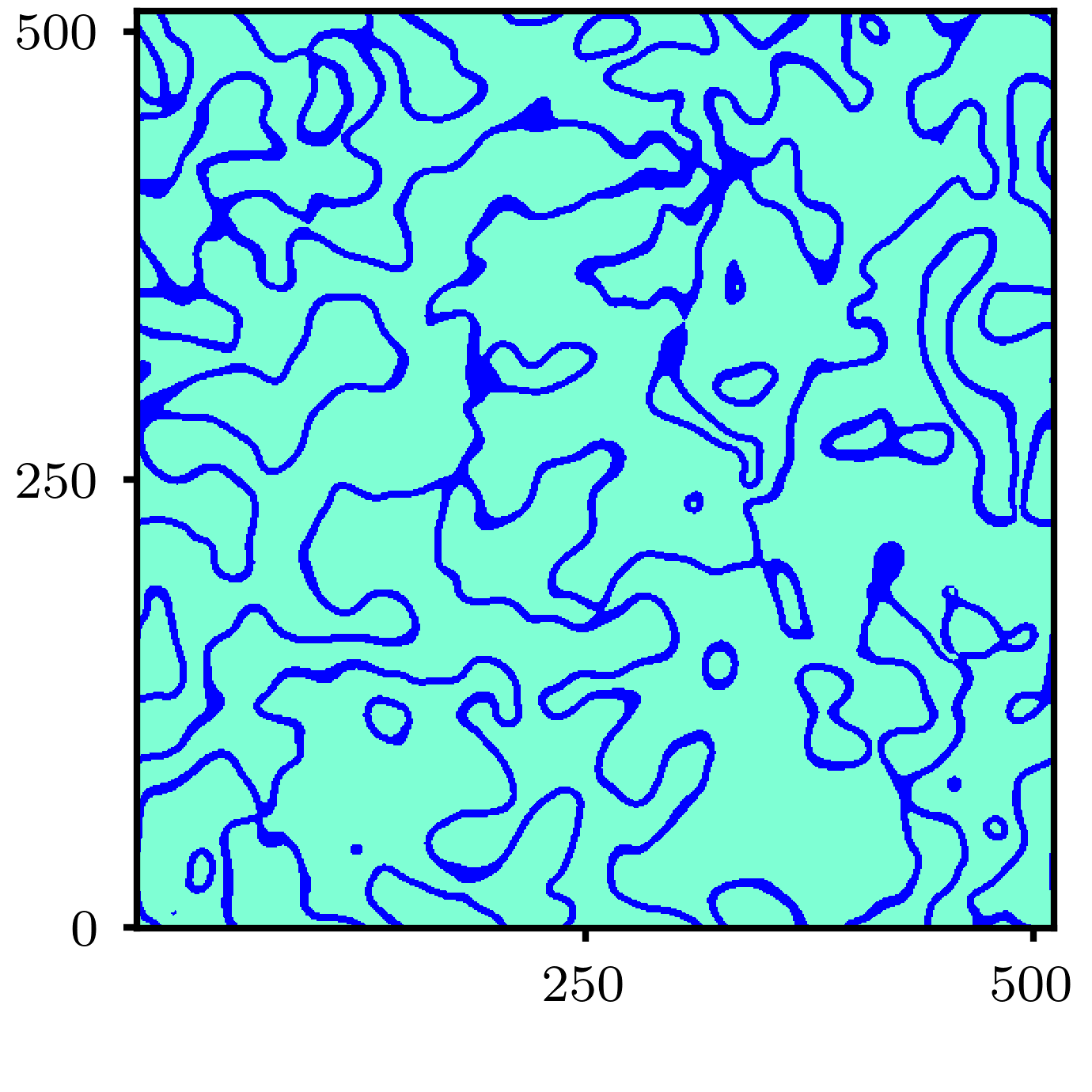}
  \subcaption{}    \label{fig:Bulk_interface_19}
     \end{subfigure}
     \hfill
     \begin{subfigure}[b]{0.66\textwidth}
         \centering
         \includegraphics[width=0.98\textwidth, height = 6.8cm]{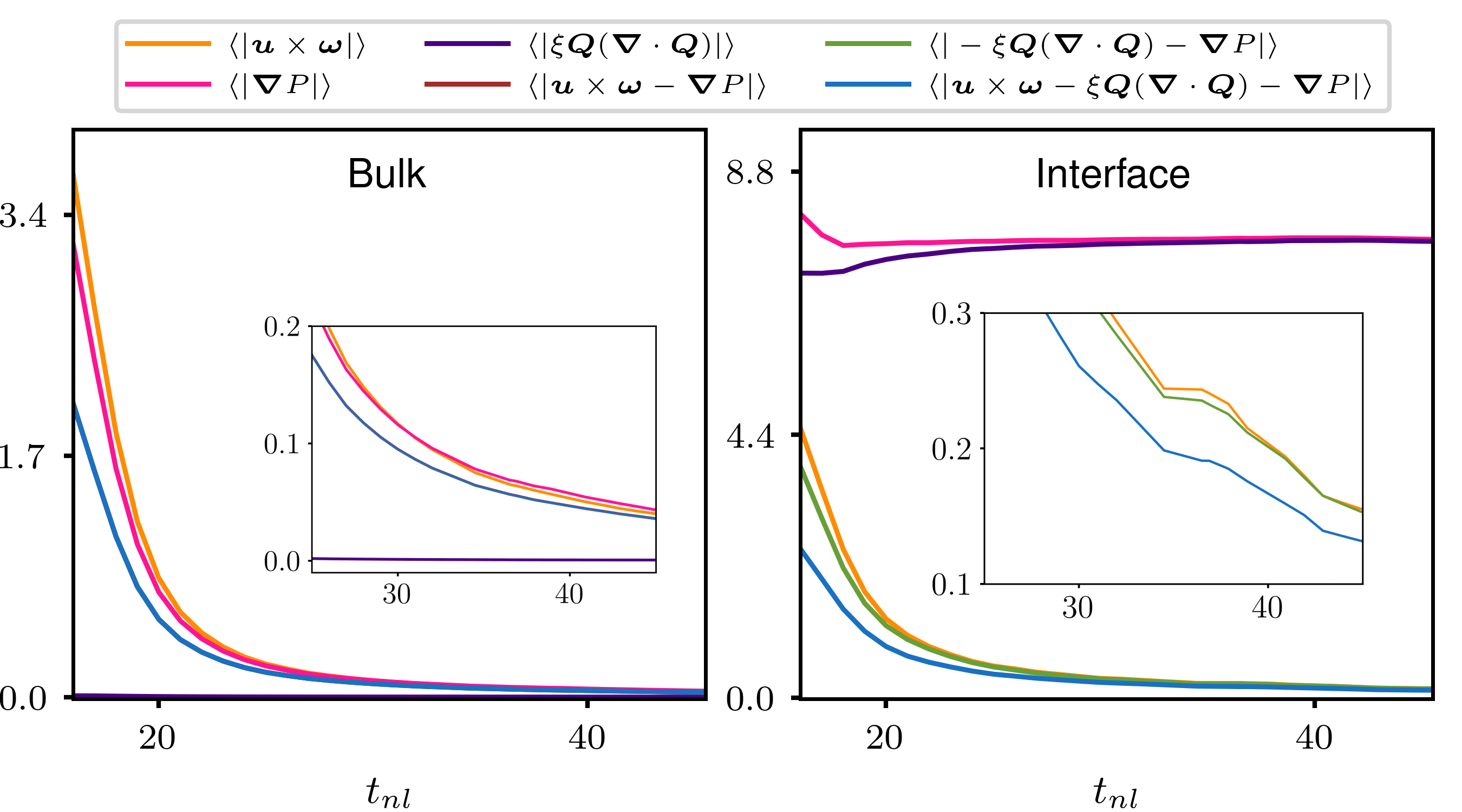}
         \subcaption{}
         \label{fig:Average_timeS}
     \end{subfigure}
        \caption{\small{\textbf{CHNS2: (\textbf{a}) Turbulent relaxation of binary fluids.} Pseudocolor plots of the composition field $\phi$. It takes values $\pm 1$ within the single fluid components which sharply changes across the interface. (left) A well-mixed phase is obtained when the turbulence is fully developed ($t_{nl} = 16$). Forcing is turned off after $t_{nl} = 16$ and the phase separation can be seen at later times (center) at $t_{nl} = 23$ and (right) at $t_{nl} = 60$, where $t_{nl}$ is the nonlinear time time. \textbf{(b) Representation of bulk and interface.} 2D slice of $|\bQ|$. Lighter regions represent bulk and darker regions represent interface (Methods). \textbf{(c) Early signs of a pressure-balanced relaxed state.} Time evolution of the average of absolute values of various nonlinear terms present in Eq.~\eqref{Eu}. 
As expected, during relaxation the nonlinear terms $\langle|\bu \times\bomega - \bnabla P|\rangle$ and $\langle|\bu \times\bomega - \xi \bQ(\bnabla \cdot \bQ) - \bnabla P|\rangle$ tend to zero at large nonlinear times in bulk and interface respectively. Note that for bulk $\langle|\bu \times\bomega - \bnabla P|\rangle$ completely overlap with $\langle|\bu \times\bomega - \xi \bQ(\bnabla \cdot \bQ) - \bnabla P|\rangle$ as the term $\bxiQdivQ \sim 0$. 
        $\langle|\bu \times\bomega|\rangle$ decays in both the regions, however, in the interfacial region, $\langle| \bnabla P|\rangle$ balances $\langle|\xi \bQ(\bnabla \cdot \bQ) |\rangle$, thereby indicating a pressure-balanced relaxed state. The average is taken over the single-fluid bulk points and two-fluid interface points separately.}}
        \label{fig:three graphs}
\end{figure*}

\section*{The CHNS model}
\label{model_invariants}
As mentioned in the introduction, the evolution of binary fluids is described by the CHNS equations\cite{berti2005,Pan2022}
 \begin{linenomath}
 \begin{align}
     \partial_{t} \bu &=  \lamb -   \xi \bQ (\bm{\nabla} \cdot \bQ) - \bm{\nabla}P + \nu \nabla^2 \bu + \bm{f} \label{Eu},\\
     \partial_{t} \phi &=  -\bu\cdot \bQ + \mathcal{M} \nabla^2 \mu , \label{EQ}
 \end{align}
\end{linenomath}
where $\bu (\bm{x}, t)$ is the velocity of the center of mass fluid, $\bomega = \bnabla \times \bu$ the vorticity field, $\bQ (\bm{x}, t)=\bm{\nabla}\phi$ the composition gradient field, where the composition field $\phi (\bm{x}, t)$ represents the local normalized density difference of the two fluids, $P = p + u^2/2 + \phi \mu - \Gamma$ is the total pressure, with $p$ being the fluid pressure, $\mu (=\delta \mathcal{F}/\delta \phi)$ the chemical potential derived from a free energy functional \begin{linenomath}$$\mathcal{F} = \int \left[ \frac{a}{2} \phi^2 + \frac{b}{4} \phi^4 + \xi (\bm{\nabla} \phi )^2\right]
d\tau$$ \end{linenomath} and $\Gamma$ the homogeneous part of the free energy. Here, we perform direct numerical simulations of CHNS equations (Methods). The entire analysis is based on the runs NS ($512^3$ hydrodynamic simulation
) and CHNS2 ($512^3$ binary fluid simulation).
To ensure that the system remains below the critical temperature $T_c$, we have taken $a=-b=-1$. Unlike a passive scalar field\cite{Yaglom1949}, the composition field $\phi$ provides feedback $-\bxiQdivQ$ to the Navier-Stokes equation due to the presence of two-fluid interfaces \cite{Cates2018},  where the parameter $\xi$ is associated with the width of the interface and is chosen in order to numerically resolve the same (Methods). The parameters $\nu$ and $\mathcal{M}$ are the fluid viscosity and the mobility coefficient respectively. Eqs.~\eqref{Eu} and \eqref{EQ} are supplemented by the incompressibility condition $\bnabla\cdot\bu=0$. 
\begin{figure*}[ht!]
\hfill

\begin{subfigure}[b]{\textwidth}
\centering
\includegraphics[width=0.95\linewidth]{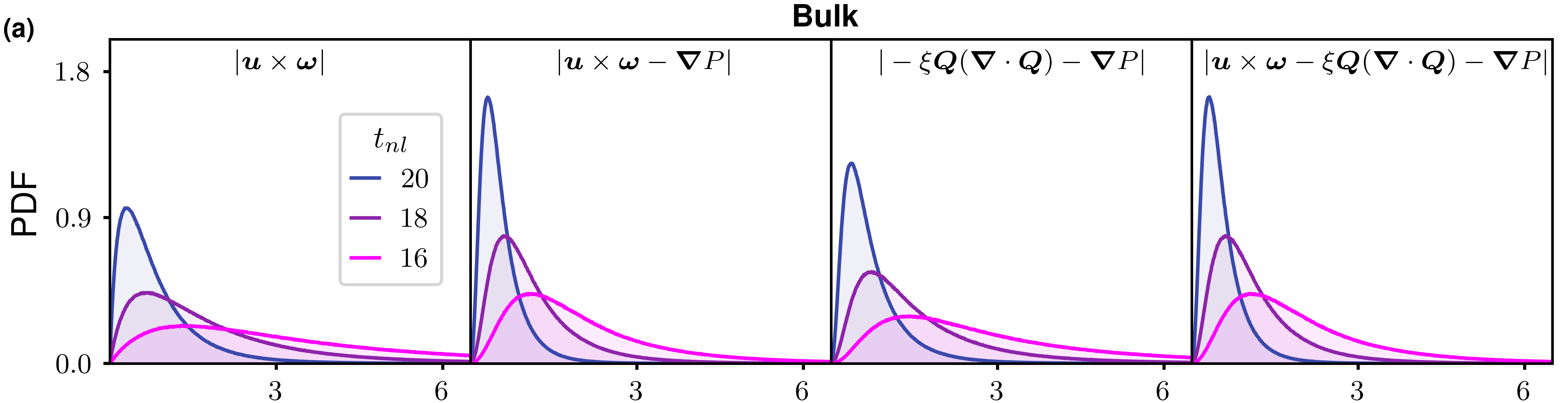} 
\phantomsubcaption{}\label{fig:Fig2_a}
\end{subfigure}\par
\begin{subfigure}{\textwidth}
\centering
\includegraphics[width=0.95\linewidth]{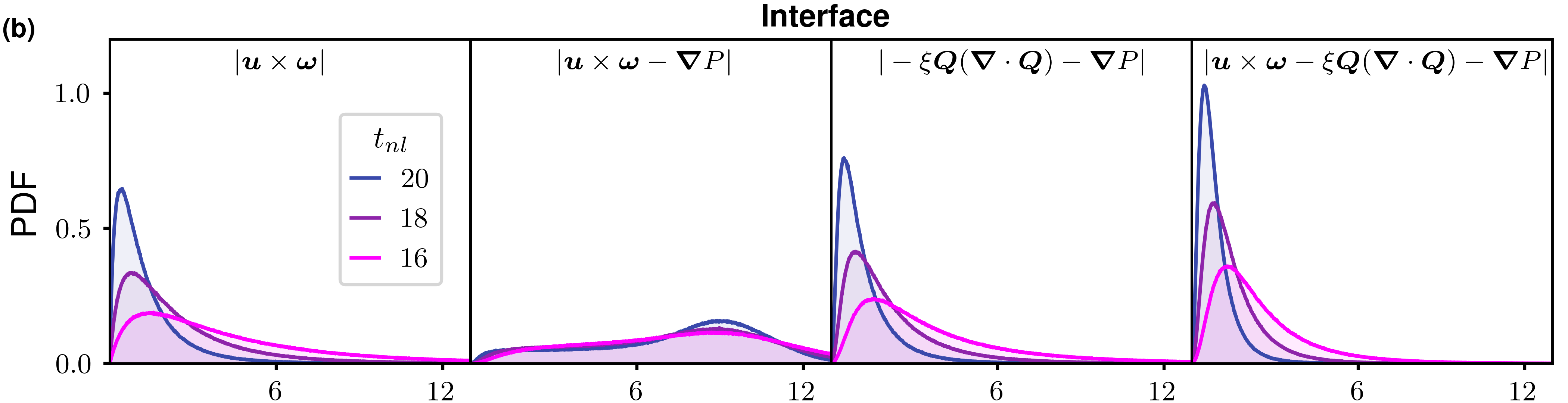}\phantomsubcaption{}\label{fig:Fig2_b}
\end{subfigure}
\caption{\small{\textbf{CHNS2: Early stage dynamics. (a), (b) } PDFs of the absolute values of various nonlinear quantities at bulk and interface respectively at different nonlinear times ($t_{nl}$) just after the initiation of relaxation. Note that, the total number of points in the bulk region increases with relaxation and that of the interface decreases. Nevertheless, having the highest peak value, the total nonlinear term $|\lamb - \bxiQdivQ-\bnabla P|$ remains the fastest decaying. }}
\end{figure*}

The large-scale stirring force $\bm{f}$ is applied to yield fully developed turbulence which subsequently leads to a phase-arrested steady emulsion state (left panel of Fig.\ref{fig:phi_color}). Upon quenching the forcing ($\bm{f} = \bm{0}$), the emulsion undergoes turbulent relaxation by domain coarsening (middle and right panel of Fig.\ref{fig:phi_color}) to ultimately attain an equilibrium phase-separated state. The whole binary fluid system can be decomposed into bulk and interface according to whether $|\bQ|=0$ or $|\bQ| \neq 0$ respectively (Fig.\ref{fig:Bulk_interface_19}). In the following, we shall systematically study and characterize the turbulent relaxation of both bulk and interface separately.

\section*{Relaxation of bulk and interface}
A system is expected to reduce its nonlinearities during turbulent relaxation \cite{Kraichnan1988, Stribling1991, Servidio2008, Matthaeus2008}. For a hydrodynamic (HD) fluid ($\bQ=\bm{0}$), the relaxed state is obtained by vanishing the net nonlinear contribution in the Navier-Stokes (NS) equations to have an alignment between $\lamb$ and $\bm{\nabla} P$ rather than a pure Beltrami-Taylor type $\bu$-$\bomega$ alignment\cite{Kraichnan1988}. For a binary fluid, we have additional nonlinear feedback $\xi\bQ\left(\bm{\nabla}\cdot\bQ\right)$ in the momentum evolution equation and it is therefore natural to study the evolution of the individual nonlinear contribution and the combination of them during the relaxation. We start by looking into the time evolutions of the averaged nonlinear terms for both the bulk and the interface before and after the quenching of the driving force in Fig.\ref{fig:Average_timeS}.  

In the bulk, we found that all $\langle|\lamb|\rangle$, $\langle|\bnabla P|\rangle$  and $\langle|\lamb-\bnabla P|\rangle$  damp out during relaxation. In addition, $\bQ$ vanishes identically in the bulk region thus leading to zero contribution from the nonlinear term $\bxiQdivQ$ to the corresponding relaxed states. In the interfacial region, again $\langle|\lamb|\rangle$ decays to zero at large times while $\bQ$ does not vanish due to the presence of a chemical potential gradient across the interface. However, unlike the bulk, $\langle|\bnabla P|\rangle$ does not fall off during relaxation. 
In particular, $\langle|\xi\bQ(\bnabla\cdot\bQ)|\rangle$ and $\langle|\bnabla P|\rangle$ are found to attain the same value at large times with $\langle|-\xi\bQ(\bnabla\cdot\bQ)-\bnabla P|\rangle$ dying out quickly indicating a balance between the feedback term and the total pressure gradient. 
Despite the apparent discrepancy in the turbulent relaxation of the bulk and the interface, it is clearly observed that the quantity $\langle|\lamb-\bxiQdivQ-\bm{\nabla}P|\rangle$
decreases to zero both for the bulk and the interfacial relaxation. In the following, we shall probe this particular aspect in more details to universally characterize the turbulent relaxation of both bulk and interface. 
In order to analyze and compare the relaxation-decay of different quantities, we study the evolution of the probability density function (PDF) of different nonlinear contributions at earlier stages of relaxation ($t_{nl} = 16, 18$ and $20$ and the forcing is withdrawn just after $t_{nl} = 16$) where the decay rates were pronounced. 
Such a study convincingly characterizes the relaxed states at every point of the flow field in terms of the local field variables. 

\begin{figure*}[ht!]
\hfill
\begin{subfigure}{\textwidth}
\centering
\includegraphics[width=0.88\linewidth]{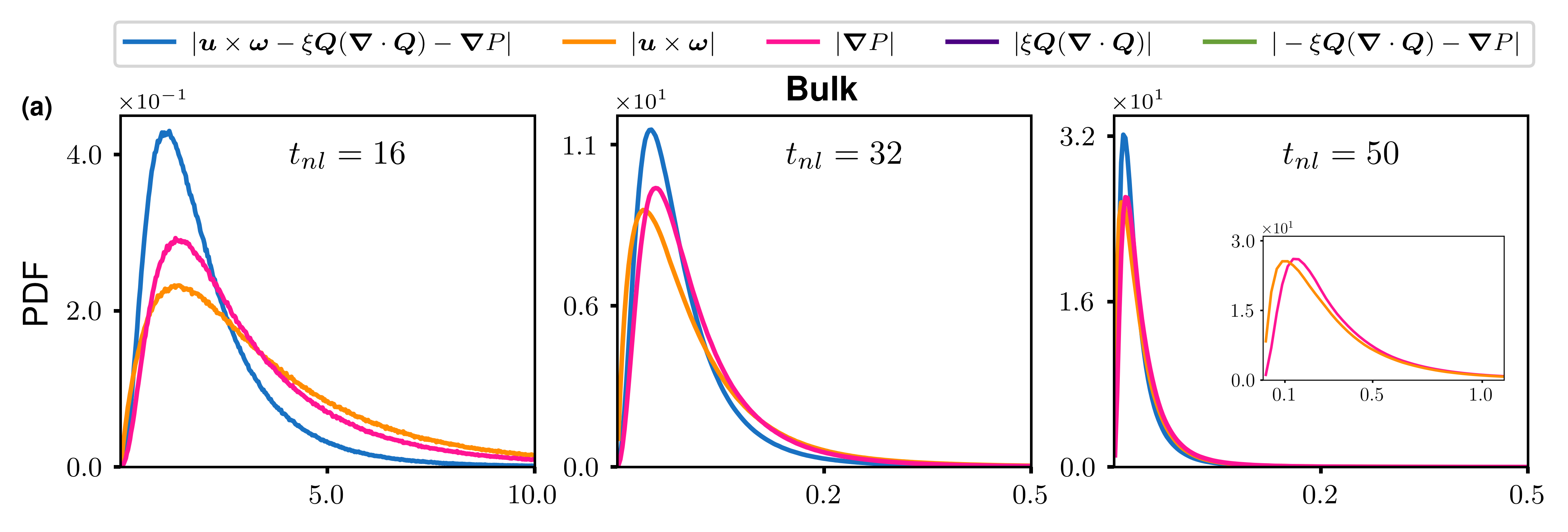}
\phantomsubcaption{}\label{fig:Fig3_a}
\end{subfigure}
\begin{subfigure}{\textwidth}
\centering
\includegraphics[width=0.87\linewidth]{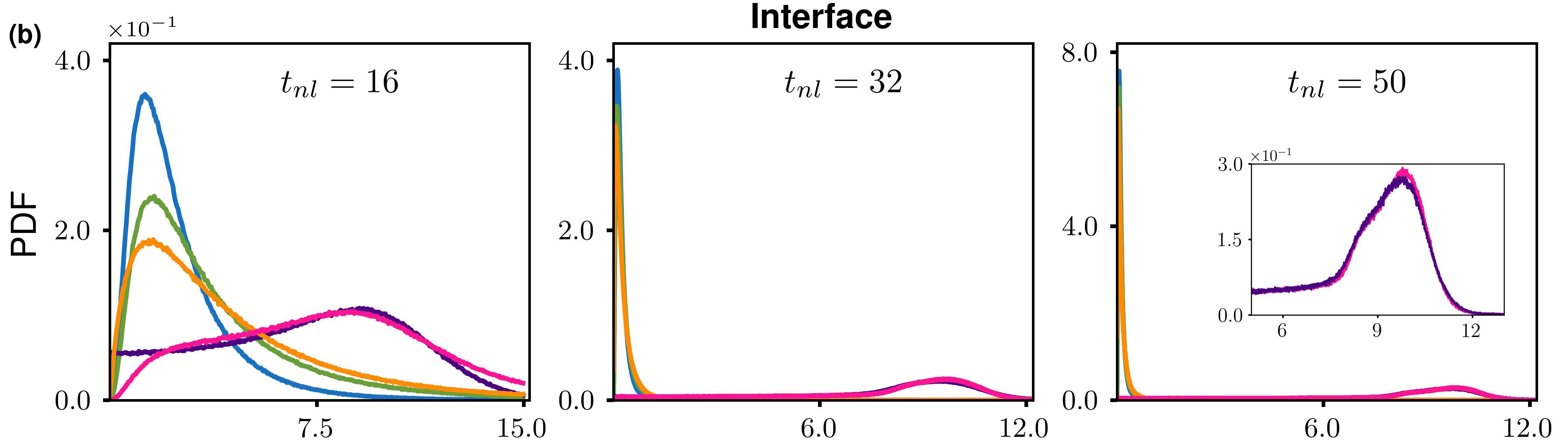}\phantomsubcaption{}\label{fig:Fig3_b}
\end{subfigure}
\caption{\small{\textbf{CHNS2:} \textbf{Late-stage dynamics.} As observed in the early stages of relaxation, the fastest decay of the total nonlinear term $|\bu \times \bomega - \xi \bQ (\bm{\nabla}\cdot\bQ)-  \bm{\nabla}P|$ ($\sim  |\bu \times \bomega -   \bm{\nabla}P|$ inside bulk) continues in the later stages as well. \textbf{(a)} Interestingly, the pressure-balanced condition of bulk ceases to hold as a $\lamb$ and $\bnabla P$ differ significantly even at later times (inset plot at $t_{nl}=50$). \textbf{(b) Pressure-balanced relaxation of the interface.} On the contrary, a clear alignment of $\bxiQdivQ$ and $\bnabla P$ can be seen on the interface.}}\label{fig:Fig3}
\end{figure*}

From Fig.\ref{fig:Fig2_a}, looking into the early stage dynamics of bulk relaxation, we observe that the peaks of the PDFs of all $|\lamb|$, $|\bnabla P| (\equiv |-\bxiQdivQ - \bnabla P|$ in bulk) and $|\lamb - \bnabla P|$ shift towards zero during the relaxation. The dominant decay of the total nonlinear term ($|\lamb - \bnabla P|$ ) also persists in the later stages of relaxation (as it is clear for $t_{nl} = 32$ and $50 $ in Fig.\ref{fig:Fig3_a}). This indicates a plausible balance between $\lamb$ and $\bnabla P$ as observed in the turbulent relaxation of an ordinary HD fluid \cite{Kraichnan1988}. However, a scrupulous investigation indicates a slight discrepancy between the PDFs of $|\lamb|$ and $|\bnabla P|$ which will be addressed later. 

For the interfacial region, a similar early-stage analysis reveals that the peak of the PDF of $|\lamb|$ shifts towards zero with time whereas no such tendency is observed for $|\lamb - \bnabla P|$ (Fig.~\ref{fig:Fig2_b}). Instead of the balance between $\lamb$ and $\bnabla P$, a clear signature of balance between $-\bxiQdivQ$ and $\bnabla P$ is found from the PDF of $|-\bxiQdivQ - \bnabla P|$. This is completely in accordance with the behaviour of the average values of absolute pressure gradient and $|\bxiQdivQ|$ in Fig.\ref{fig:Average_timeS}.

In later times, this balance becomes more prominent, e.g., at $t_{nl} = 50$, PDFs of $|\bnabla P|$ and $|\bxiQdivQ|$ completely overlap thereby confirming strong equality between the two vectors at every  point on the interface (Fig.\ref{fig:Fig3_b}). The aforesaid equality thus indicates a pressure-balanced relaxed state on the interface. In addition, both $|\lamb|$ and $|- \bxiQdivQ - \bnabla P|$ die off during the relaxation and so does the total nonlinear term $|\lamb - \bxiQdivQ - \bnabla P|$. In fact, similar to the bulk, the total nonlinear term remains the fastest decaying quantity (light blue curves in Fig.\ref{fig:Fig3_a} and \ref{fig:Fig3_b}), thus providing a universal characterization of the turbulent relaxation of a binary mixture.
\begin{figure}[ht!]
    \centering
    \includegraphics[width=0.35\textwidth]{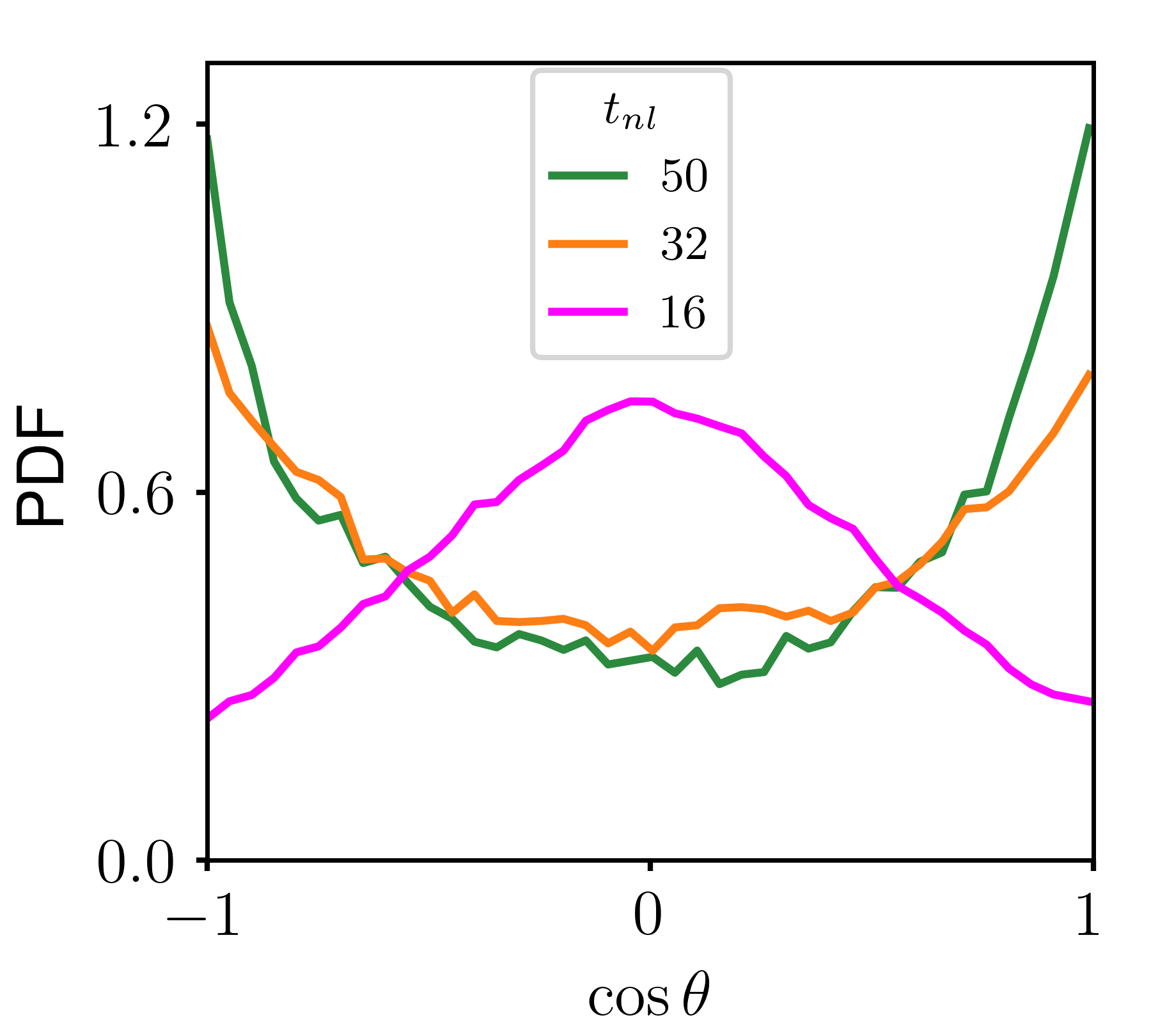}
  \caption{\small{\textbf{CHNS2: Quiescent condition for interfacial relaxation.} PDF of $\cos \theta$, where $\cos \theta =  \frac{\bm{u}\cdot \bm{Q}}{|\bm{u}||\bm{Q}|}$. The broadening of the PDF towards $\pm 1$ with time clearly signifies a lack of perpendicularity between the two, justifying $\bu=\bm{0}$.}}   \label{fig:uQ}
  \end{figure}

\begin{figure*}
 \centering
    \includegraphics[width=0.98\textwidth, ]{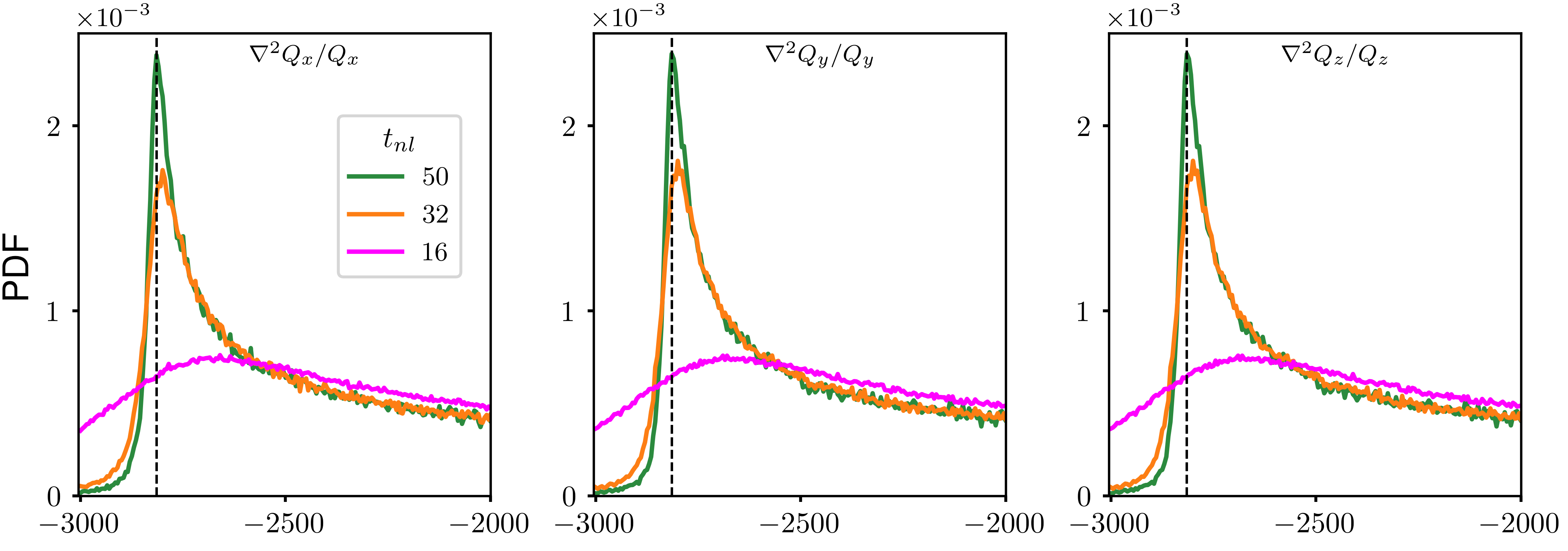}
\caption{\small{\textbf{CHNS2: Helmholtz-like relaxation equation for the interface through the pressure-balanced condition.} The PDFs of all three components of $\nabla^2\bQ/\bQ$ sharply peak towards the same value during relaxation, thus verifying Eq.~\eqref{Helmholtz} with $\lambda \simeq -2815$ (dashed line in all three panels).}} 
 \label{fig:Helmholtz_condition}
\end{figure*}

\section*{Linking bulk and interface relaxation: a universal theory}
Despite markedly distinct relaxation processes, the bulk and the interfacial relaxations are characterized by the suppression of a common decaying nonlinear quantity. In the following, we provide a theoretical justification for such universal relaxation using PVNLT\cite{Banerjee2023}. According to this principle, the relaxed states are obtained when the average scale-to-scale transfer of a cascading invariant identically vanishes at each scale within the inertial range\footnote{The range of length scales free from macroscopic and microscopic effects and is principally governed by the nonlinearity.}.

Incompressible CHNS system allows two ideal invariants- (i) the total energy $E \ [=\int  (u^2 + \xi Q^2)/2 \ d\tau]$ and (ii) the scalar energy $S \ (= \int \phi^2/2\ d \tau)$. Inside the inertial range, the average scale-to-scale transfer of $E$ and $S$ can be written in terms of two-point ($\bx$ and $\bx'$) correlators  as\cite{Pan2022} 
\begin{linenomath}
\begin{align}
    \langle\mathcal{F}_{tr}^E\rangle
&= \left\langle \bu^{\prime} \cdot \left[  \bu \times \bomega - \xi \bQ (\bm{\nabla}\cdot\bQ) - \bm{\nabla} P \right]\right. \nonumber\\
&\left.+
 \bu\cdot [ \bu^{\prime} \times \bomega^{\prime} - \xi \bQ^{\prime}(\bm{\nabla}^{\prime}\cdot\bQ^\prime) -\bm{\nabla}^{\prime} P^{\prime}]\right.\nonumber\\  
 &\left.- \xi \bQ^{\prime}\cdot\bm{\nabla}(\bu\cdot\bQ)-\xi\bQ \cdot\bm{\nabla}^{\prime}(\bu^{\prime} \cdot \bQ^{\prime})\right\rangle,\\
 \langle\mathcal{F}_{tr}^S\rangle &=  \langle \phi^{\prime} (\bu\cdot\bQ) + \phi ( \bu^{\prime}\cdot\bQ^{\prime}) \rangle,\label{Ftrs_Sp}
\end{align}
\end{linenomath}
where $\langle\cdot\rangle$ denotes the ensemble average (equivalent to space average in homogeneous turbulence). Relaxation essentially implies the vanishing of $\langle\mathcal{F}_{tr}^E\rangle$ and $\langle\mathcal{F}_{tr}^S\rangle$ at each scale of the inertial range. For the case where $\bu$ may or may not be zero, the vanishing of $\langle\mathcal{F}_{tr}^E\rangle$ provides
\begin{linenomath}
\begin{align}
\bu \times \bomega - \xi \bQ (\bm{\nabla}\cdot\bQ)-  \bm{\nabla}P &=  \bm{\nabla} \Phi, \label{PVNLT_Ep} \\
\bm{\nabla}(\bu \cdot \bQ ) &= \bnabla\times\bm{A} \label{PVNLT_Ep1},
\end{align}
\end{linenomath}
where $\Phi$ and $\bm{A}$ are an arbitrary scalar and vector fields respectively.
Similarly, for vanishing $\langle\mathcal{F}_{tr}^S\rangle$ and non-vanishing $\phi$, one obtains 
\begin{linenomath}
\begin{equation}
\hspace{2.5cm}    \bu \cdot \bQ = 0 \label{PVNLT_Sp1}.
\end{equation}
\end{linenomath}
 
Inside the bulk $\bQ = \bm{0}$ and hence Eq.~\eqref{PVNLT_Sp1} is trivially satisfied similar to Eq.~\eqref{PVNLT_Ep1}, if $\bm{A}$ is chosen to be irrotational. On the interface, where none of $\bu$ and $\bQ$ vanishes, Eqs.~\eqref{PVNLT_Ep1} and \eqref{PVNLT_Sp1} require point-wise perpendicularity of $\bu$ and $\bQ$. However, the distribution of the angle between $\bu$ and $\bQ$ shows a broader spread instead of a sharp peak at $90^{\circ}$ (Fig.\ref{fig:uQ}). This apparent puzzle can be reconciled by means of a careful inspection of \eqref{Ftrs_Sp} which shows $\langle\mathcal{F}_{tr}^S\rangle = - \langle \phi^{\prime} \bu^{\prime} \cdot\bQ + \phi  \bu\cdot\bQ^{\prime} \rangle$ for homogeneous turbulence.
For $\langle\mathcal{F}_{tr}^S\rangle$ to vanish at all scales inside the inertial range, at least one of $\bu$ and $\bQ$ must vanish.
Relaxation of the interface therefore necessarily requires $\bu$ to vanish identically.

Choosing $\bnabla \Phi = \bm{0}$ in Eq.~\eqref{PVNLT_Ep}, one recovers the universal relaxation of both the bulk and the interface as described in Fig.\ref{fig:Fig3}. In particular, 
\begin{linenomath}
\begin{align}
\hspace{-2cm}\text{for bulk ($\bQ =\bm{0}$)} &: \,\, \bu\times \bomega = \bnabla P  
    \quad \text{and}  \label{Bulkp} \\
\hspace{-0.5cm}    \text{for relaxed interface ($\bu =\bm{0} $)} &:\,\,
\xi \bQ (\bm{\nabla}\cdot\bQ) =  -\bm{\nabla}P. \label{PVNLT_Interface2}
\end{align}
\end{linenomath}
By taking the curl on both sides of Eq.~\eqref{PVNLT_Interface2} and using $\bnabla \times \bQ = \bm{0}$, we obtain the following Helmholtz-like condition
\begin{linenomath}
\begin{equation}
\bQ \times \bnabla (\bnabla \cdot \bQ) = \bm{0} \implies \nabla^2 \bQ = \lambda \bQ\implies \nabla^2\phi=\lambda\phi + C, \label{Helmholtz}
\end{equation}
\end{linenomath}
where $\lambda$ and $C$ are arbitrary constants. 
This relation is clearly verified in Fig.\ref{fig:Helmholtz_condition} whence the value of $\lambda$ is found to be nearly $-2815$.

 \begin{figure*}[ht!]
\centering
    \includegraphics[width=0.7\textwidth]{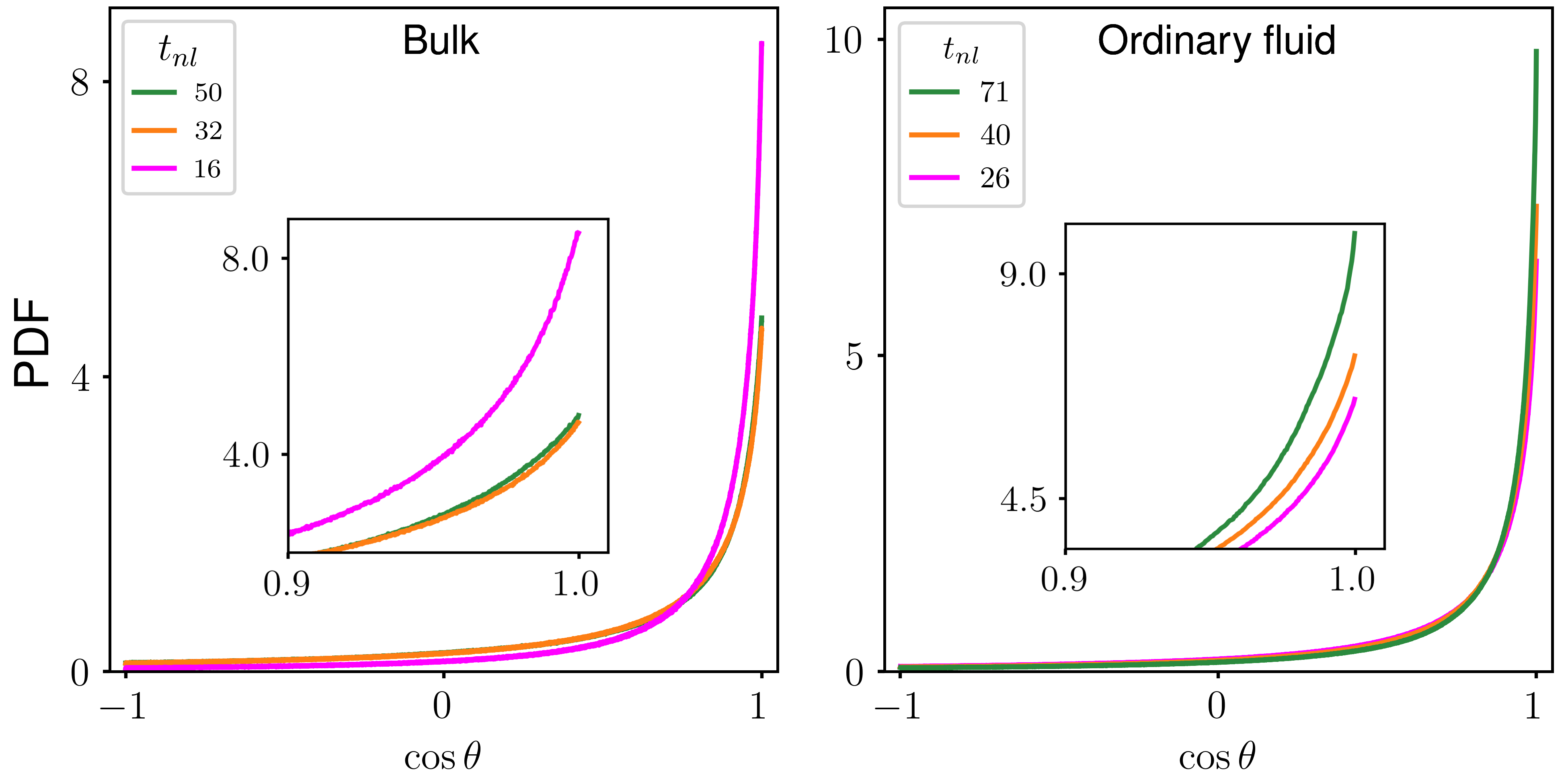} \caption{\small{\textbf{CHNS2 and NS: Atypical relaxation of bulk}. The cosine of the angles between $\lamb$ and $\bnabla P$. The pressure-balanced state gets weaker for bulk relaxation (left) with time in contrast with the corresponding HD fluid (right). Thus the bulk relaxation condition \eqref{Bulkp} will hold trivially for zero flow. Note that for HD, the forcing is switched off after $t_{nl} = 26$.}}  
\label{fig:Bulk_HD}
\end{figure*}
\section*{Bulk relaxation vs. single fluid relaxation} 
A close inspection of the Fig.\ref{fig:Fig3_a} (third panel) shows that
although the peak of the PDF of $|\lamb - \bnabla P|$ shifts rapidly towards zero during relaxation, a non-negligible gap is visible between the individual PDFs for $|\lamb|$ and $|\bnabla P|$  even at large times. The possibility of a non-trivial balance between the two can therefore be discarded. In addition, the alignment between the two is also found to reduce significantly during the relaxation (Fig.\ref{fig:Bulk_HD}). The only possibility that the Eq.~\eqref{Bulkp} holds, therefore, requires both sides of the equation to vanish identically at relaxation. This is in direct contrast with what is observed for an ordinary fluid, where the alignment between $\lamb$ and $\bnabla P$ increases in the course of relaxation (Fig.\ref{fig:Bulk_HD}). 
Interestingly, for the interface, a non-trivial balance between the feedback term and the total pressure gradient is meticulously obeyed as the PDFs of $|\bxiQdivQ|$ and $|\bnabla P|$ are found to overlap in Fig.\ref{fig:Fig3_b} (inset of the third panel). 

\section*{Discussion}
In this work, we have systematically investigated the turbulent relaxation of a binary fluid and prescribed a universal way to describe both the bulk and the interfacial relaxation. We have found a clear signature of a pressure-balanced relaxed state for the interface whereas the bulk is found to relax only when it attains a static state. This is a stark difference with respect to the turbulent relaxation of an ordinary hydrodynamic fluid where a balance between pressure gradient and the nonlinear term takes place before the fluid comes to rest. For a binary fluid system, this discrepancy can be attributed to the conservation of scalar energy which identically vanishes at every point of a single fluid. Despite this difference, the relaxation of both the bulk and the interface can be interpreted as a tendency of suppression of the total nonlinear term $\lamb - \bxiQdivQ - \bnabla P$.

In addition, the pressure-balanced relaxation on the interface can be reduced to a Helmholtz-like condition as it is seen in Eq.~\eqref{Helmholtz}. This kind 
of topographic relaxation is also observed in various other systems e.g. strongly rotating two-dimensional flows \cite{bretherton1976}, gravitational relaxation of mantle diapirs of Venus \cite{Janes1995}, formation of anti-cyclonic eddies over oceanic basins \cite{solodoch2021} and in particular, where the flow is governed by a scalar field e.g., the stream function in 2D HD and the magnitude of the magnetic vector potential in 2D MHD turbulence\cite{Banerjee2023}.

The present study entirely implements the principle of vanishing nonlinear transfer to find the relaxed states of a turbulent binary mixture under critical temperature. Note that, the principle of selective decay cannot give the correct relaxed states as obtained in Eqs.~\eqref{Bulkp} and \eqref{PVNLT_Interface2} as, by construction, it cannot capture a finite pressure gradient in the turbulent relaxed states. 

The current 
study can be extended to investigate turbulent relaxation of more complex binary fluid systems $e.g.,$ biological systems such as dilute bacterial suspensions\cite{tiribocchi2015}, phytoplankton suspensions\cite{peters2000} and synthetic active colloids\cite{campbell2019}, etc.,  which exerts extra feedback to the Navier-stokes equation owing to its activity \cite{Pan2022, Cates2018}. Broadly, this will open up future research possibilities of controlling and engineering different active systems from biologically and chemically relevant perspectives.


\section*{Methods}

\subsection*{Simulation details}
Eqs.~\eqref{Eu} and \eqref{EQ} are simulated using pseudo-spectral method in three dimensions in a 2$\pi$-periodic box with $N$ grid points in each direction. Due to the presence of a cubic non-linearity in chemical potential, a $N/2$-dealiasing method is employed \cite{canuto2007spectral,Orszag_1971}.
\begin{table}[ht]\label{Table1}
\centering
\begin{tabular}{|l|l|l|l|l|l|}
\hline
 & $N$ & $\nu (\times 10 ^{-3})$ & $\xi (\times 10 ^{-3})$ & $\mathcal{M}$ & $Re$ \\
\hline
NS & $512$  & $0.6$ & - & - & 3000\\
\hline
CHNS1 & $256$  & $2$ & $1.0$ & 0.01 & 453\\
\hline
CHNS2 & $512$ & $0.6$ & $0.3$  & 0.01 & 1225 \\
\hline
\end{tabular}
\caption{\label{tab} NS and CHNS stand for single-fluid Navier-Stokes and Cahn-Hiliard-Navier-Stokes simulations respectively.}
\end{table}
The system is initialized below $T_c$ from rest ($\bu = \bm{0})$ in a phase-mixed state with a uniform distribution of $\phi$ with $-0.05\leq\phi(\bm{x},0)\leq 0.05$. The simulation code is parallelized using a python-based message-passing interface (MPI) scheme\cite{Mortensen2016HighPerformance}.

In the absence of any forcing, the two fluids will eventually phase-separate in regions with $\phi = \pm 1$ (determined from the free energy functional). 
\begin{figure}
\includegraphics[width=0.9\linewidth]{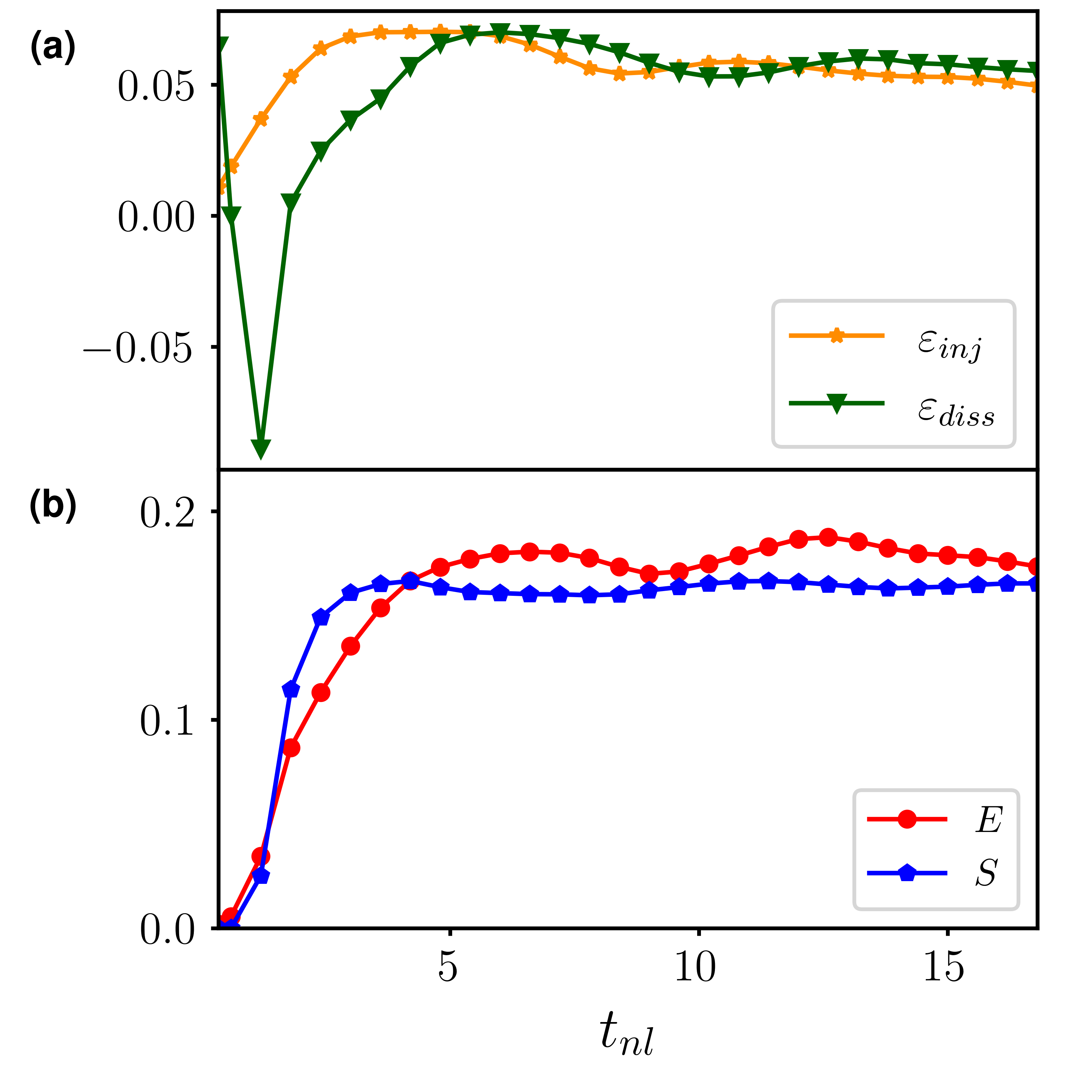}
    \caption{\small{ \textbf{CHNS2}:
    \textbf{Conservation of inviscid invariants: (a)} Average injection $\varepsilon_{inj}\ (= \langle \bm{f}\cdot \bm{u} \rangle)$ balances the average dissipation $\varepsilon_{diss} (= \langle \nu \omega^2 - \xi \mathcal{M} \bQ\cdot \nabla^2\bnabla \mu \rangle)$ \textbf{(b)} leading to the time-invariance of $E$ and $S$.} }
    \label{fig:diss}
\end{figure}
The arrest of phase separation is achieved through turbulence by stirring the momentum evolution equation \eqref{Eu} with a large-scale non-helical forcing $\bm{f}$,
\begin{linenomath}
\begin{align}
 \bm{f}= &f_0 sin(k_f x)cos(k_f y)cos(k_f z)\hat{x}\nonumber \\ 
 &- f_0 cos(k_f x )sin(k_f y)cos(k_f z)\hat{y},\nonumber 
\end{align}
\end{linenomath}
where $k_f$ and $f_0$ are the forcing  wavenumber and amplitude respectively. For the current study, we choose $k_f = 2$, $f_0 = 0.5$. The typical domain size of such a turbulent phase-arrested binary fluid system is determined by the Hinze-criterion \cite{Hinze1955}. The system is evolved in time with an unconditionally stable  explicit-implicit scheme \cite{Eyre1998, Yoon2020, Li2021}, until it reaches a statistical steady state  (Fig.\ref{fig:diss}). The parameter $\xi $ is proportional to the square of the width of the two-fluid interface. Ideally, the interface is very sharp $i.e.,$ the width of the interface (the length over which $\phi$ changes from 0.9 to -0.9) is negligibly small. However, to have a numerically well-resolved interface, it is chosen to cover at least six grid points \cite{berti2005,perlekar2019}. Other parameters $\nu$ and $\mathcal{M}$ are set in order to have a reasonably high Reynolds number ($R_e$) turbulence (Table \ref{tab}).

As an initial step, we performed the simulation with $256^3$ grid points with appropriate parameters given in Table \ref{tab}. Main findings of our study is found to remain unchanged with respect to the analysis carried out with $512^3$ grid points (mentioned in the main text). For the ordinary HD fluid simulation, we employ the same non-helical forcing $\bm{f}$ and a standard fourth order Runge-Kutta scheme for time-integration.
\subsection*{Characterization of bulk and interface}
The whole binary fluid has been divided into into bulk and interfacial regions. Ideally, $\bm{Q}$ should be zero within bulk as it consists only of single fluid phase of the two-fluid mixture. However, due to numerical limitations and for computational convenience, we choose $|\bm{Q}|\leq 1.2$ for bulk region. Similarly, the interface is a sharp-transition region between the two-fluid ($|\bm{Q}|>>1)$), however for our study the interface is defined for $|\bm{Q}|\geq 21$ (Fig.\ref{fig:Bulk_interface_19}). Although $\bQ$ is not identically zero inside bulk, it does not effect the analysis as the value of $-\bxiQdivQ$ is found to be negligibly small (Fig.\ref{fig:Average_timeS}).  
\subsection*{Probability density functions (PDFs)}
Below the critical temperature, binary fluid always minimizes the interfacial regions in order to minimize the free energy $\mathcal{F}$. During relaxation, this leads to the increase and decrease of bulk and interface points respectively.
Note that, here a standard definition of PDF has been employed where the number of counts in each bin has been divided by the total count and the corresponding bin size. The simulations have been carried out upto $t_{nl} = 60$ (Fig.\ref{fig:phi_color}) and the tendency of the universal relaxation remains unchanged as seen from the Fig.\ref{fig:Bulk_inter_timS} below. However, for the sake of visual clarity, the PDFs in the main text are plotted for three different nonlinear times upto $t_{nl} = 50$. 
\begin{figure}[ht!]
\centering
\includegraphics[width=0.45\textwidth]{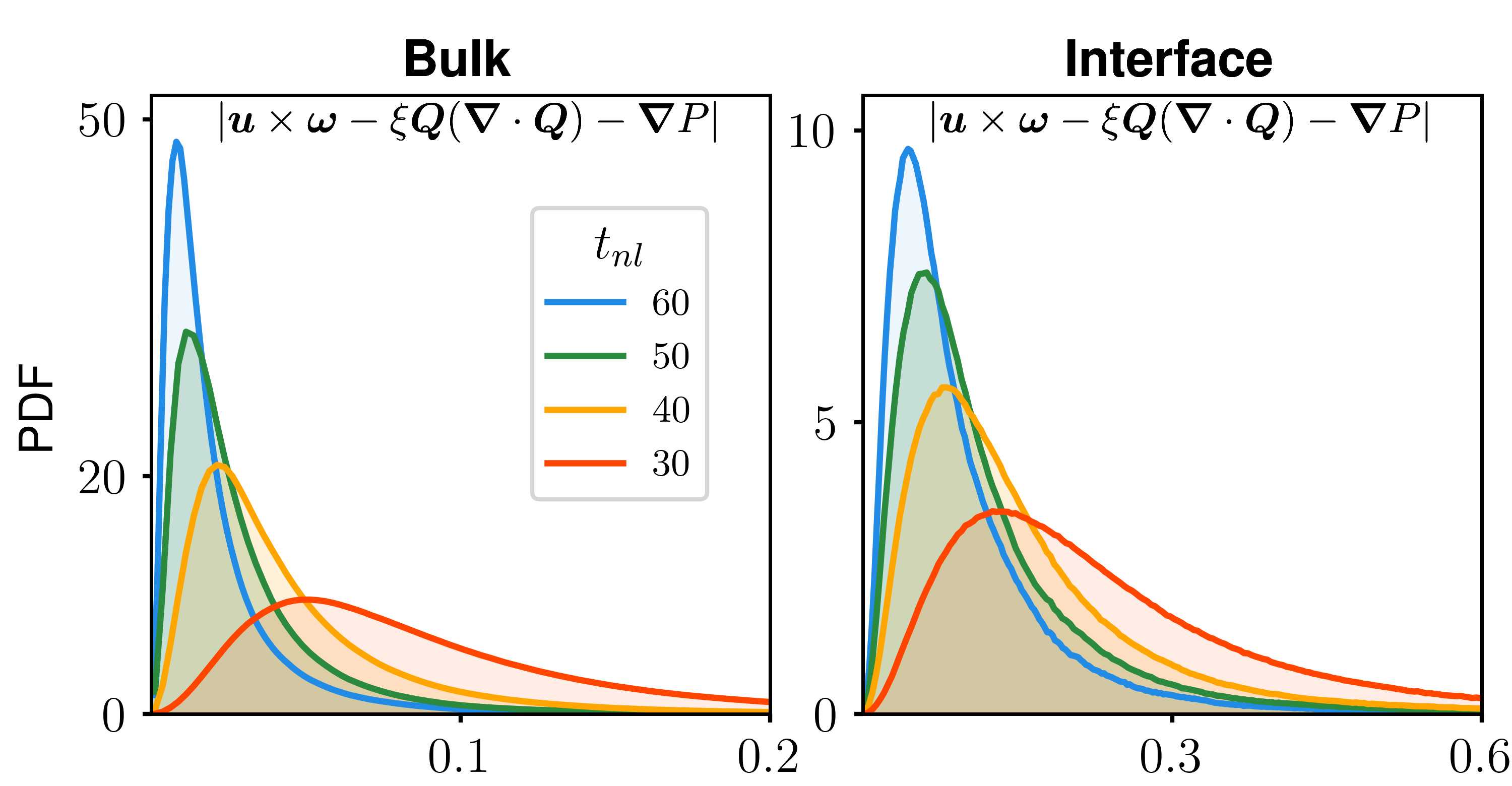} \caption{\small{\textbf{CHNS2: Universal relaxation of bulk and interface.} The universal nature of binary fluid relaxation is captured by the suppression of $|\lamb-\bxiQdivQ-\bnabla P|$ which decays quicker than other nonlinear contributions both for the bulk and the interface. This fact is further theoretically supported by the principle of vanishing nonlinear transfer.}}  
\label{fig:Bulk_inter_timS}
\end{figure}

\section*{Acknowledgements}
The authors acknowledge useful discussions with Anando G. Chatterjee. The simulation code is developed by the first author following the parallelization scheme provided in ref \cite{Mortensen2016HighPerformance}. The simulations are performed using the support and resources provided by PARAM Sanganak under the National Supercomputing Mission, Government of India at the Indian Institute of Technology, Kanpur. S.B. acknowledges DST-INSPIRE faculty Research Grant No. DST/PHY/2017514 and CEFIPRA Research Grant No. 6104-01. 
\section*{Author contributions}
S.B. and N.P. conceived the idea and designed the study. N.P. performed the numerical simulation. S.B., N.P. and A.H. analysed and interpreted the results and contributed to the composition of the manuscript.

\end{document}